\documentclass[9pt,twocolumn,twoside]{opticajnl}
\journal{opticajournal} 

\setboolean{shortarticle}{false}

\usepackage{lineno}

\title{Quantum eraser enables single-shot quantum holography with undetected photons}

\author[1,2,3,*]{Naru Yoneda}
\author[1,3]{Nao Sakiyama}
\author[2]{Luis Ordóñez}
\author[1]{Vishal Prajapati}
\author[1]{Kanako Yoshimura}
\author[1,2]{Osamu Matoba}

\affil[1]{Graduate School of System Informatics, Department of System Science, Kobe University, Rokkodai 1-1, 5 Nada, Kobe 657-8501, Japan}
\affil[2]{Center for Life Photonic Innovation (CLiPI), Kobe University, Rokkodai 1-1, Nada, Kobe 657-8501, Japan}
\affil[3]{These authors contributed equally to this work.}

\affil[*]{yoneda.naru@port.kobe-u.ac.jp}

\begin{abstract}
Quantum holography with undetected photons (QHUP) enables the complex amplitude of infrared to be acquired beyond the sensitivity range of silicon based sensors, through induced coherence withoud induced emission.
QHUP generally requires multiple phase-shifted holograms to retrieve the complex-amplitude distribution of an object, which limits measurement stability and temporal resolution.
Although classical parallel phase-shifting holography provides a favorable balance between temporal and spatial resolution, it has not previously been implemented in QHUP because of the constraints imposed by photon distinguishability.
In this study, we propose a quantum eraser-based geometric phase-shifting QHUP.
By accurately erasing not only polarization distinguishability but also temporal distinguishability arising from differences in the optical paths, the complex-amplitude distribution of an object can be retrieved in a single shot using a silicon-based image sensor equipped with a polarization-filter array.
This single-shot acquisition, combined with real-time processing, enables the observation of dynamic objects in the near-infrared spectral region. 
To the best of our knowledge, these experimental results constitute the first real-time acquisition of the complex-amplitude distribution of a near-infrared object using QHUP with a silicon-based image sensor equipped with a polarization-filter array. 
Furthermore, to enhance the reproducibility and applicability of this study, the experimental optical setup, bill of materials, and source code for real-time imaging have been made openly available.
The ability to retrieve the complex-amplitude information carried by undetected photons in real time is expected to extend the applicability of QHUP to a wide range of fields, including spectroscopic imaging and semiconductor inspection.
\end{abstract}

\setboolean{displaycopyright}{false} 

\begin{document}

\maketitle

\section{Introduction}
Optical technologies beyond the capabilities of classical optics can be realized using quantum optics such as single-photon sources, squeezed light, and entangled photon pairs \cite{Moreau2019,Defienne2024}.
In particular, quantum imaging based on entangled photons has enabled highly sensitive imaging \cite{Mitchell2004,Ono2013,Black:23}, super-resolution imaging \cite{He2023}, ghost imaging \cite{PhysRevA.52.R3429,PhysRevLett.74.3600,Moreau:18}, and noise-robust imaging \cite{doi:10.1126/sciadv.aax0307, 10.1063/5.0314450, doi:10.1126/sciadv.aay2652}.
Among these techniques, quantum imaging with undetected photons (QIUP) enables information about an object to be obtained without the object's interaction with photons being directly detected \cite{Lemos2014,BarretoLemos:22,https://doi.org/10.1002/qute.202300353}. 
This principle is based on induced coherence without induced emission \cite{PhysRevLett.67.318} and can exploit frequency-entangled photon pairs.
Consequently, QIUP can retrieve information in the infrared spectral region, which cannot be directly detected using conventional silicon-based image sensors. 
Furthermore, the information obtained in QIUP also contains the effects experienced by the visible photons. 
Therefore, for the undetected infrared photons, optical components designed for visible wavelengths can be utilized not only for detection but also for modulation \cite{Li:18,Wolley2026}.
Its potential applications include adaptive optics \cite{Li:18}, pupil plane modulation \cite{Wolley2026}, gas sensing \cite{10.1063/5.0242197,neves2026}, IR polarimetry \cite{Paterova:19,Chakraborty:25}, video-rate imaging \cite{https://doi.org/10.1002/lpor.202000327}, spectroscopy \cite{Kalashnikov2016,Tashima:24}, and optical coherence tomography \cite{Paterova_2018,Vanselow:20}.
\begin{figure*}[t]
\centering
\includegraphics[width=\linewidth]{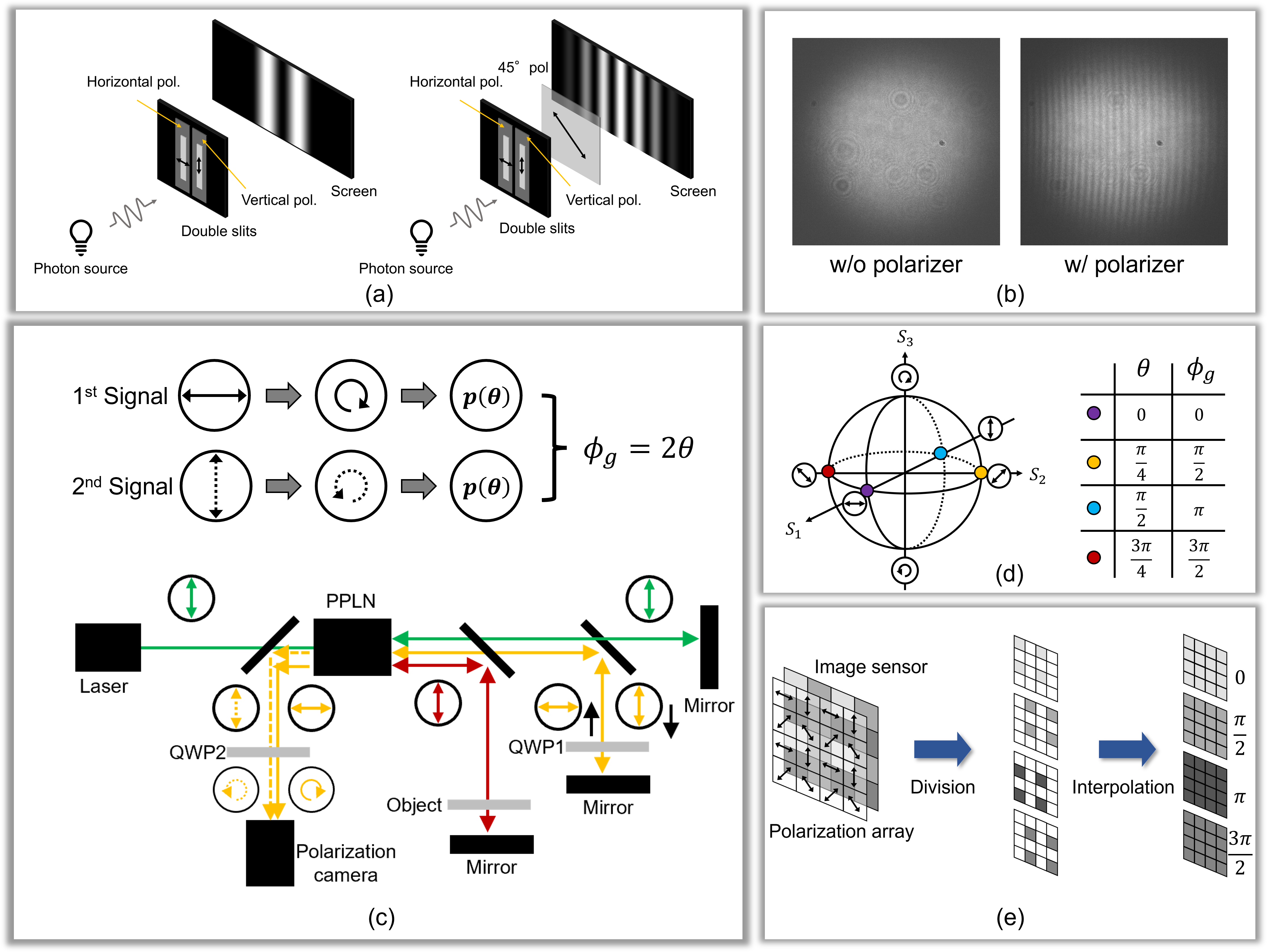}
\caption{Conceptual diagrams of the key components of the proposed method.
(a) Schematic of the basic quantum-eraser concept, in which the polarizer eliminates which-path distinguishability.
(b) Results of the quantum eraser in the proposed method: left, without the polarizer; right, with the polarizer inserted.
(c) Trajectories of the polarization states in the optical system; the solid line denotes the first SPDC process, and the dashed line denotes the second SPDC process.
(d) Illustration of the geometric phase shifts in the proposed method and the Poincaré sphere.
(e) Conceptual diagram of the parallel phase-shifting method using a camera equipped with a polarization-filter array.
}
\label{fig:QHUP_Fig.1}
\end{figure*}
\begin{figure*}[t]
\centering
\includegraphics[width=160mm]{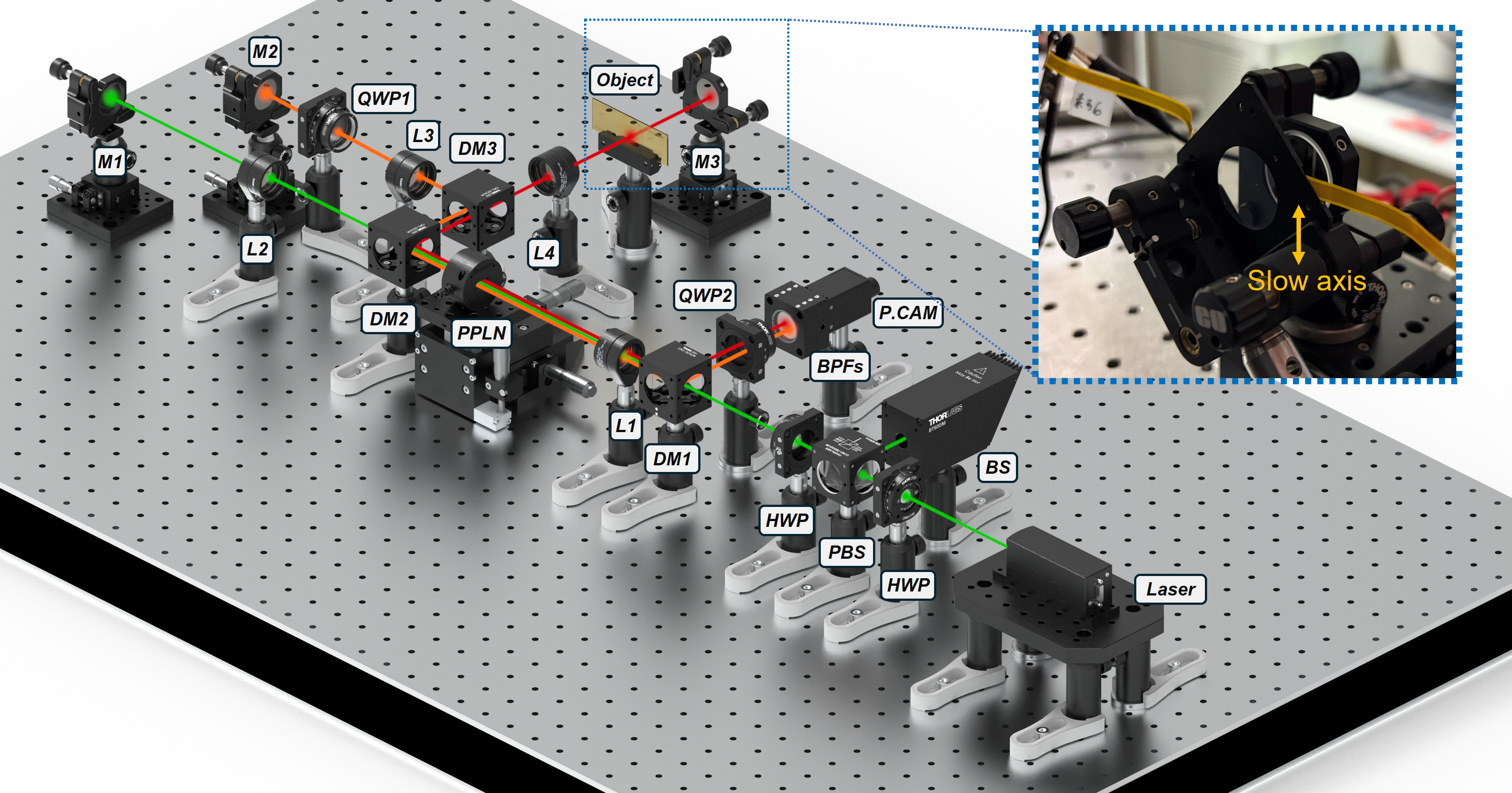}
\caption{Experimental setup of the proposed method. The green, orange, and red rays represent the 532 nm pump, 810 nm signal, and 1550 nm idler beams, respectively. HWP: Half-wave plate, QWP: quarter-wave plate, PBS: polarization beam splitter, DM: dichroic mirror, M: mirror, L: lens, BPF: bandpass filter, BS: beam stop. A photograph of the LC lens used in the experiment is shown in the upper-right corner.
}
\label{fig:QHUP_Fig.2}
\end{figure*}
Properties of QIUP, such as resolution and magnification, have already been evaluated \cite{Hochrainer:17,Fuenzalida2022resolutionofquantum}.
Although QIUP can obtain only amplitude information of a target interacted with undetected photons, quantum holography with undetected photons (QHUP) enables the reconstruction of both the amplitude and phase distributions of an object in the infrared region \cite{doi:10.1126/sciadv.abl4301}. 
However, multiple interference patterns with different bias phases are generally required to retrieve the complex-amplitude distribution of the object. 
To reduce the number of required measurements, the off-axis method based on Fourier fringe analysis has been proposed \cite{Leon-Torres:24,Pearce2024,Topfer:25,https://doi.org/10.1002/qute.70355}. 
This approach improves the temporal resolution of QHUP and enables hologram acquisition at the frame rate of the image sensor.
However, this is achieved at the cost of reduced spatial-frequency bandwidth and, consequently, lower resolution.
As an alternative, a parallel phase-shifting interferometric method using a phase-shift array displayed on a spatial light modulator has also been reported \cite{Topfer:25,Li:18}.
This method can reduce the spatial-resolution limitation of the off-axis configuration.
However, the use of a spatial light modulator increases the complexity of the optical system. 
In addition, the use of spatial light modulators results in unwanted higher-order diffraction, which limits their diffraction efficiency and thus reduces the overall light utilization efficiency.
A phase-quadrature technique has also been proposed, which enables simultaneous acquisition of the interference patterns required for phase shifting \cite{Haase:23}.
This method involves dividing the imaging sensor into four regions and acquiring interference fringes with different phase shifts in each region.
However, this approach suffers from a relatively complex optical configuration and difficulty in selecting appropriate regions of the recorded interference patterns for extraction.
In conventional holography, parallel phase-shifting methods \cite{10.1063/1.1777796} using polarization cameras and geometric phase shifts have been proposed to acquire the multiple phase-shifted holograms required for phase retrieval in a single shot \cite{Tahara:11,Tahara_2017,Choi:18,Yoneda:23,Nobukawa:26}. These methods have enabled the measurement of complex-amplitude distributions of dynamic objects. However, parallel phase-shifting using a polarization camera has not yet been demonstrated in QHUP. This is because the use of geometric phase shifting in QHUP requires careful control of the distinguishability of the photons contributing to the interference, a requirement that has not been sufficiently addressed in previous studies.

In this study, we propose a single-shot QHUP method based on geometric phase shifting with a polarization camera, eliminating the need for multiple image acquisitions. 
In the proposed method, the distinguishability between photon pairs in QHUP is deliberately erased through polarization modulation.
The resulting geometric phase shifts are observed as interference fringes through a quantum-erasure process. 
A CMOS camera equipped with a polarization-filter array acts as the quantum eraser, allowing interference patterns with different geometric phase shifts to be recorded simultaneously at different pixels. 
Therefore, by applying a reconstruction algorithm used in conventional parallel phase-shifting digital holography, the complex-amplitude distribution of an object in the infrared region can be retrieved in a single shot.

\section{Principle}
QHUP exploits the indistinguishability of the SPDC fields generated during the first and second passes.
This allows information about an object in the idler path to be detected through the signal path.
When the SPDC fields generated during the two passes have mutually orthogonal polarizations, the two generation paths become distinguishable.
This situation is analogous to the which-path-marked double-slit experiment shown in Fig. \ref{fig:QHUP_Fig.1}(a). As a result, no interference fringes can be observed.
In contrast, as shown on the right-hand side of Fig. \ref{fig:QHUP_Fig.1}(a), inserting a polarized erases the which-path information and restores the interference fringes \cite{PhysRevA.25.2208}.
A similar phenomenon can also be observed in QHUP by carefully considering polarization and temporal distinguishability. 
Figure \ref{fig:QHUP_Fig.1}(b) shows the results of a quantum-eraser experiment in QHUP (Details of the optical experiment are provided in the Supplemental Section 1). 
When the signal photons generated during the first and second SPDC processes have mutually orthogonal polarizations, no interference fringes are observed [left part of Fig. \ref{fig:QHUP_Fig.1}(b)]. 
In contrast, inserting a polarizer in front of the sensor erases the polarization-based distinguishability between the two signal fields, thereby restoring the interference fringes [right part of Fig. \ref{fig:QHUP_Fig.1}(b)].

When the SPDC fields generated during the first and second passes undergo different polarization transformations, the resulting contribution appears as a geometric phase \cite{Pancharatnam1956,10.1098/rspa.1984.0023}.
The phase difference between the two signal fields can be decomposed into dynamic and geometric contributions. 
The dynamic phase originates from the optical path-length difference, whereas the geometric phase arises from the polarization evolution.
The signal-photon detection probability, $P_s$, can then be expressed as:
\begin{equation}
P_s(\theta)
=
|g|^2
\left[
1+
\sqrt{\eta_o}
\cos
\left(
\Delta\Phi_0
+
\phi_o
+
\phi_g
\right)
\right].
\label{eq:1}
\end{equation}
\begin{figure*}[t]
\centering
\includegraphics[width=\linewidth]{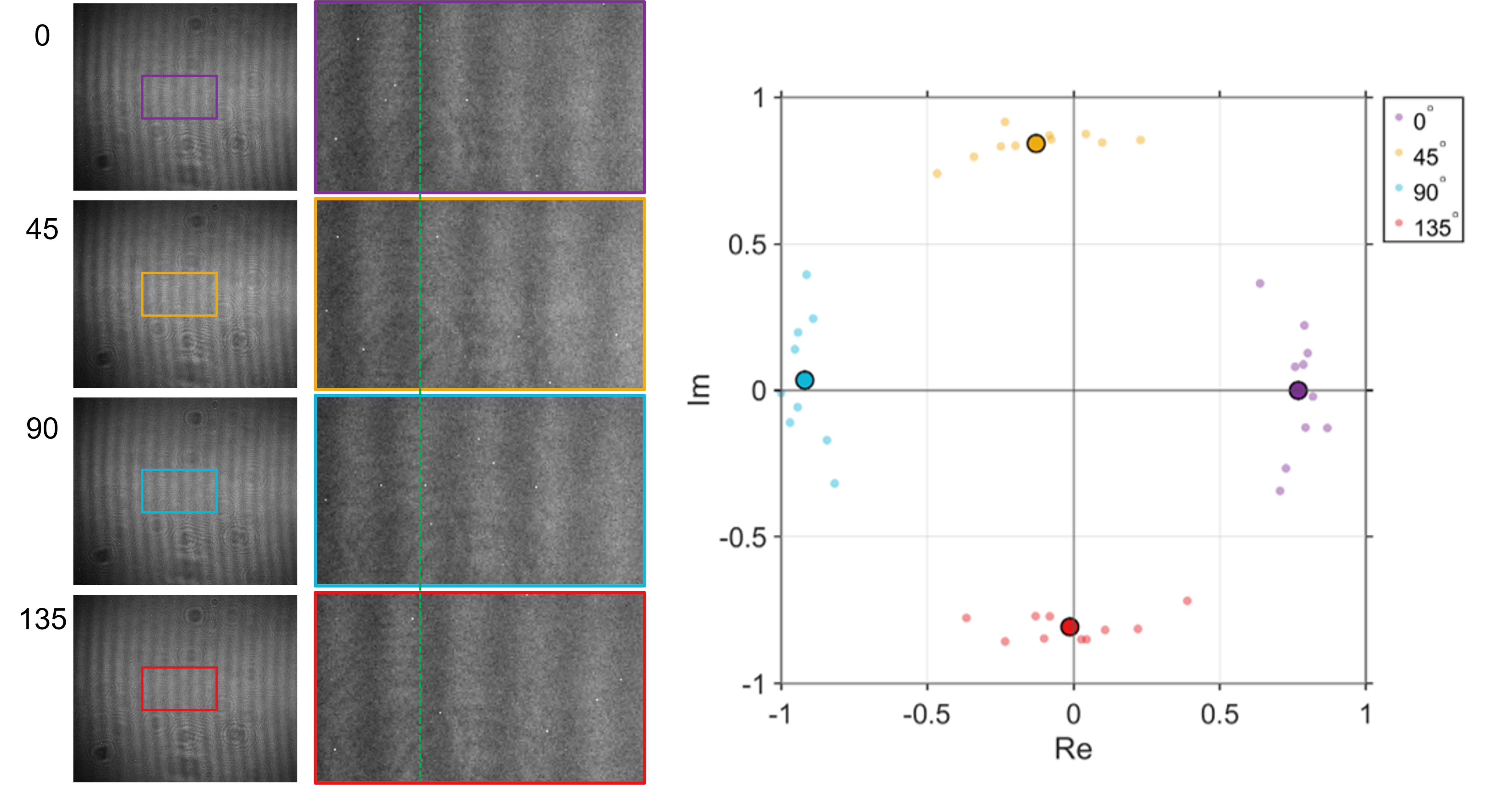}
\caption{Experimental results for the measurement of geometric phase shifts. The holograms shown in purple, yellow, cyan, and red correspond to polarizer angles of \(0^\circ\), \(45^\circ\), \(90^\circ\), and \(135^\circ\), respectively. Enlarged views of the central regions are shown to the right of each hologram. The green dashed lines indicate reference positions and clearly visualize the phase shifts of the interference fringes. The constellation map on the right quantitatively evaluates the geometric phase shifts. The smaller points represent phase-shift values measured at different times, while the larger markers indicate their mean values.
}
\label{fig:QHUP_Fig.3}
\end{figure*}
\begin{figure}[t]
\centering
\includegraphics[width=\linewidth]{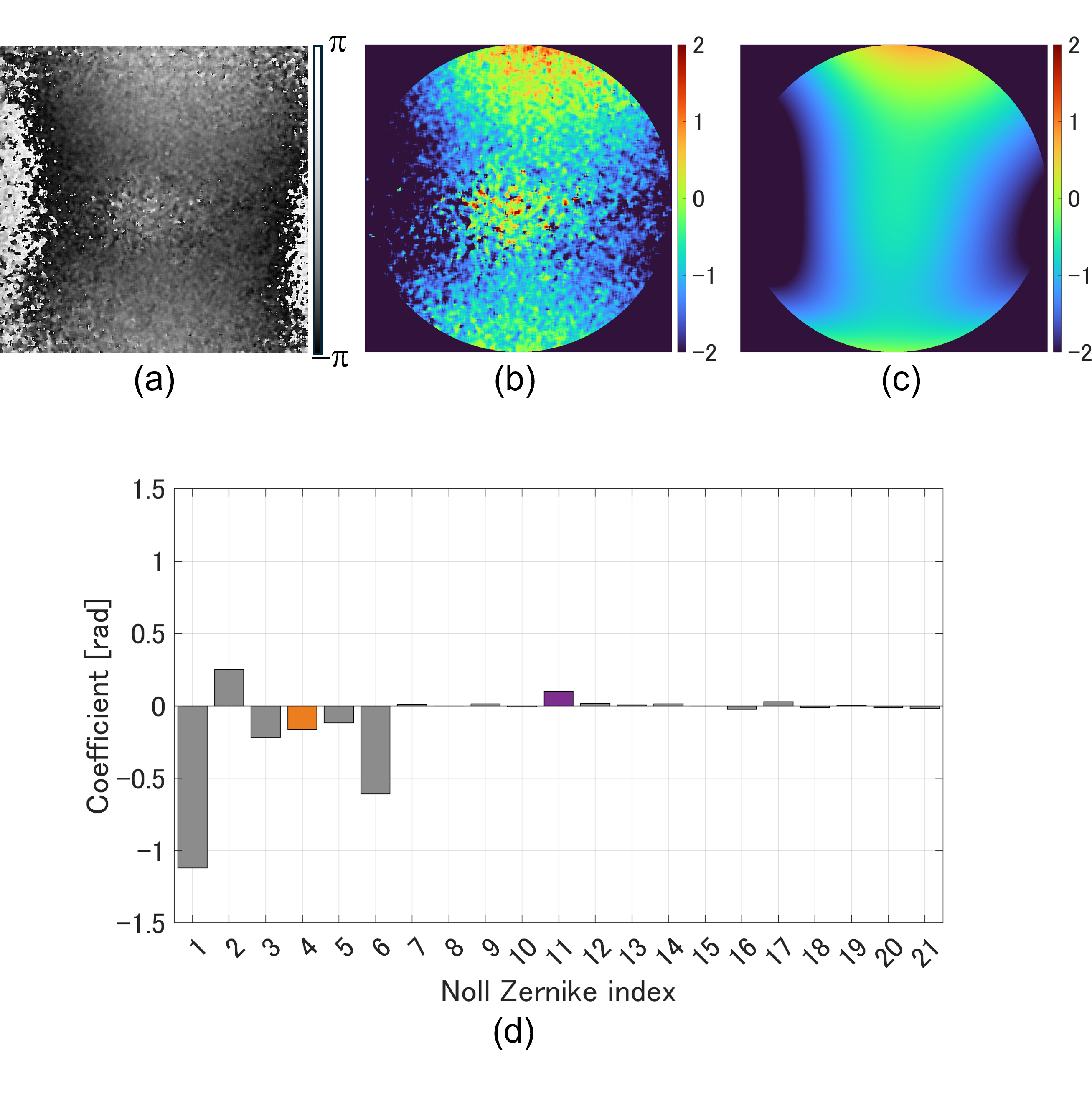}
\caption{Aberration measurement results. Phase distribution (a) after smoothing, (b) after phase unwrapping, fitted with Zernike polynomials, and (d) corresponding Zernike coefficients.
}
\label{fig:QHUP_Fig.4}
\end{figure}
where $g$, $\sqrt{\eta_o}$, $\phi_o$, and $\Delta\Phi_0$ are the amplitude of the pair-generation, the amplitude of the object, the dynamic phase of the object, and the bias phase caused by the phase difference between the first and second paths of the pump, idler, and signal photons.
In Eq. (\ref{eq:1}), $\phi_{g}$ is the geometric phase difference of the signal photons.
A detailed explanation of Eq. (\ref{eq:1}) can be seen in the Supplementary Section 2. 
In the conventional phase-shifting interferometric method \cite{Yamaguchi:97}, four step phase-shifts are typically used: 0, $\pi/2$, $\pi$, and $3\pi/2$ to acquire the dynamic phase of the object.
To generate these phase shifts geometrically in QHUP, the signal fields produced during the first and second passes are subjected to the polarization transformations illustrated in Fig. \ref{fig:QHUP_Fig.1}(c).
Here, neglecting the constant phase shifts $\Delta\Phi_0$ introduced in the signal, idler, and pump optical paths, and a geometric phase shift caused by a first signal photon experiencing a quarter-wave plate twice, the detection probability can be described as:
\begin{equation}
P_s(\theta)
=
|g|^2
\left[
1+
\sqrt{\eta_o}
\cos
\left(
\phi_o
+
2\theta
\right)
\right],
\label{eq:2}
\end{equation}
where $\theta$ is the orientation angle of the polarizer.
This expression shows that interference patterns with different phase shifts can be generated by varying the angle of a polarizer.
The geometric phase is given by one-half of the solid angle enclosed by the polarization trajectory on the Poincaré sphere \cite{Jackin:16,Choi:18}.
As indicated by the final points on the Poincaré sphere in Fig. \ref{fig:QHUP_Fig.1}(d), the geometric phase shift depends on the final polarization angle.
Therefore, in the optical configuration shown in Fig. \ref{fig:QHUP_Fig.1}(c), the phase of the interference pattern can be controlled by changing the orientation of the polarizer placed in front of the camera. 
When a polarization camera, a CMOS sensor equipped with a polarization-filter array, is used, different phase-shifted interference patterns are recorded at different pixels, as shown in Fig. \ref{fig:QHUP_Fig.1}(e). By extracting the pixel values associated with each polarizer orientation and applying spatial interpolation, the set of interference patterns required for phase-shifting interferometry can be obtained simultaneously.
Thus, the polarization-filter array acts both as a quantum eraser and as a mechanism for introducing pixel-dependent geometric phase shifts.

\begin{table}[t]
\centering
\caption{Results of the relative mean geometric phase shifts.}
\label{tab:GP}
\begin{tabular}{c|cccc}
\hline
$\theta$ [rad]& 0 & $\pi/4$ & $\pi/2$ & $3\pi/4$ \\
\hline
$\phi_g'$ [\( ^\circ\)]& 0 & 98.6 & 177.8 & 269.1 \\
\hline
\end{tabular}
\end{table}
\section{Experiments}
\subsection{Setup}
The experimental optical setup is shown in Fig. \ref{fig:QHUP_Fig.2}, with additional details provided in \cite{yoneda_2026_22659491}.
We created a three-dimensional schematic of the optical setup and made it openly available, allowing readers to inspect the experimental configuration from various viewing angles \cite{yoneda_2026_22659491}.
The major optical components required for the experiment are also made openly available as a bill of materials \cite{yoneda_2026_22659491}.
The optical system without polarization equipment is configured based on the previous research \cite{Pearce:23}.
A single-mode semiconductor laser operating at a wavelength of 532 nm (J100GS11, Kyocera) is focused by Lens 1, which has a focal length of 150 mm (AC254-150-AB-ML, Thorlabs), onto a periodically poled lithium niobate (PPLN) crystal (MOPO515-0.5, Covesion). 
The PPLN crystal has a length of 10 mm and a poling period of $7.40~\mu\mathrm{m}$.
The PPLN crystal was mounted on an oven (PV10, Covesion), and its temperature was maintained at 135 °C using a temperature controller (OC3, Covesion).
The optical fields emerging from the PPLN crystal through type-0 phase matching are separated into the pump and SPDC fields by dichroic mirror DM2 (DMSP650R, Thorlabs). 
Signal and idler photons at wavelengths of 810 and 1550 nm, respectively, are generated through SPDC. 
The pump beam passes through Lens 2 with a focal length of 150 mm (LA1433-ML, Thorlabs), is reflected by Mirror 1  (BB1-E02, Thorlabs), and returns along the original optical path. 
The SPDC field reflected by DM2 is further separated into the signal and idler fields by DM3 (DMLP950R, Thorlabs). 
The idler field transmitted through DM3 undergoes complex-amplitude modulation by the object, is reflected by Mirror 3 (PF10-03-P01, Thorlabs), and returns along the original path.
The signal field reflected by DM3 passes through a quarter-wave plate (QWP1), is reflected by Mirror 2 (BBB-111-E03, Thorlabs), and propagates back along the same path. 
The slow axis of QWP1 was set at 45 degrees with respect to the horizontal axis.
Because the signal field passes through QWP1 twice, its polarization is converted into a linear polarization orthogonal to that of the signal field originally generated in the PPLN crystal.

The returning pump, signal, and idler fields then pass through the PPLN crystal. During this second passage, the pump field generates another pair of signal and idler photons through SPDC.
The optical fields subsequently pass through Lens 1, and only the SPDC fields are reflected by DM1 (DMSP650R, Thorlabs). 
After passing through QWP2, the signal fields generated during the first and second passages are converted into circularly polarized states with mutually orthogonal handedness. 
The slow axis of QWP2 was set at 45 degrees with respect to the horizontal axis.
Consequently, the signal fields generated during the first and second SPDC processes have mutually orthogonal polarizations and are therefore distinguishable, preventing the observation of interference fringes.
\begin{figure*}[t]
\centering
\includegraphics[width=\linewidth]{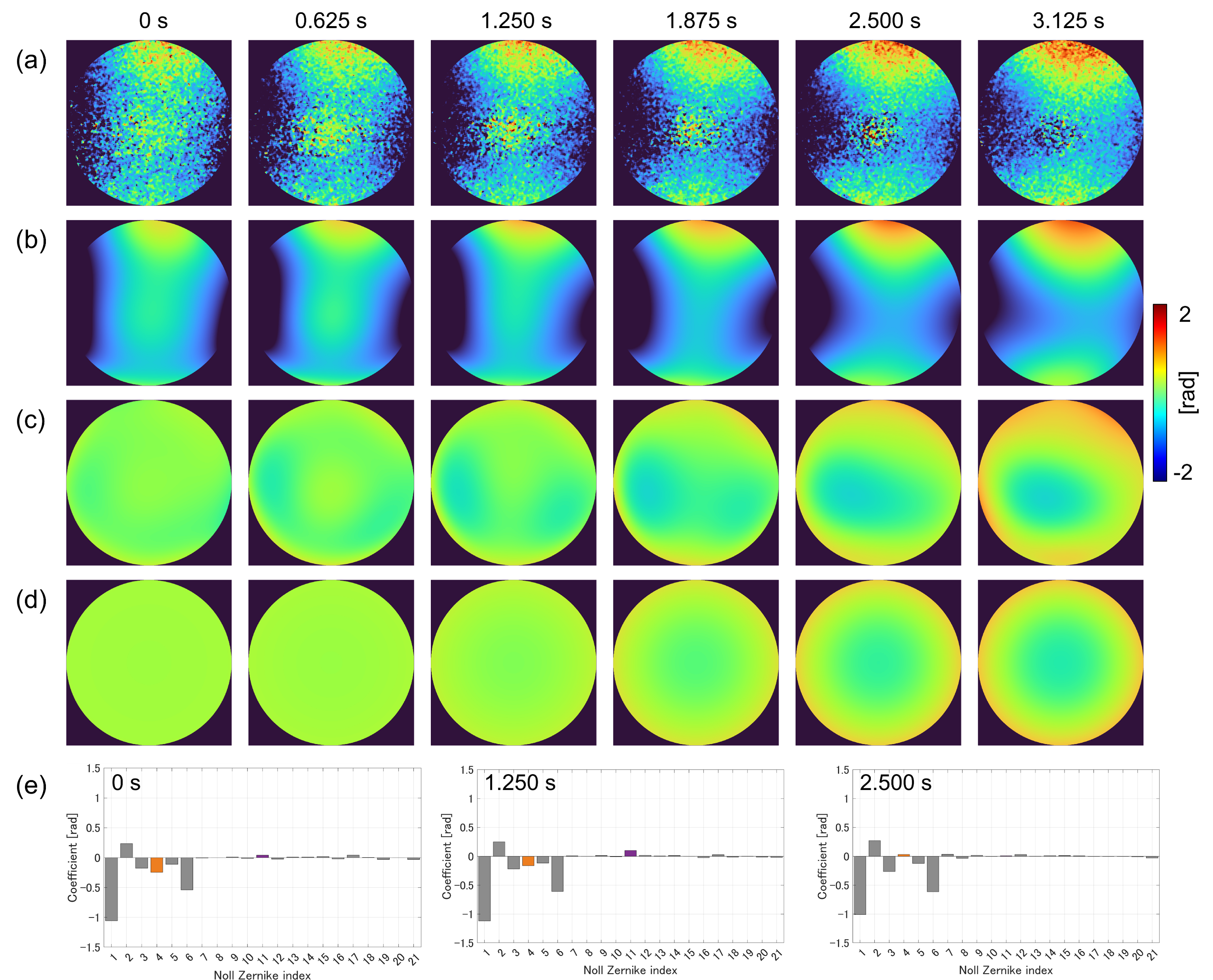}
\caption{Wavefront measurement results obtained while applying a voltage to the LC lens.
(a) Phase distributions after smoothing and phase unwrapping,
(b) after fitting with Zernike polynomials,
(c) obtained after subtracting the initial phase shown in Fig. \ref{fig:QHUP_Fig.4}, and
(d) showing only the defocus term.
(e) Temporal variations of the Zernike coefficients.
}
\label{fig:QHUP_Fig.5}
\end{figure*}

After the residual pump and idler light is removed using band-pass filters (FBH808-3, FBH800-40, Thorlabs), the signal fields are detected by a polarization camera, a CMOS sensor equipped with a polarization-filter array  (TRI050S-MC, LUCID VISION LABS).
The number of pixels and the pixel size are 2448 $\times$ 2048 and 3.45 $\mu$m.
The polarizers in the array project the two orthogonally polarized signal fields onto common polarization states, thereby erasing the polarization information that distinguishes the signal photons generated during the first and second passages. 
As a result, quantum interference fringes are observed. 
Because these fringes encode the information carried by the idler field, optical information at 1550 nm can be acquired using a CMOS camera that detects the 810 nm signal field.
As shown in Fig. \ref{fig:QHUP_Fig.1}(d), the difference in the geometric phase shifts experienced by the two signal fields depends on the transmission-axis orientation of the polarizer. 
Therefore, the four interference patterns required for a phase-shifting method can be acquired simultaneously.
The holograms were obtained by extracting the pixels corresponding to each polarizer transmission-axis orientation from the hologram acquired with the polarization camera and interpolating the extracted data using the bicubic method \cite{Xia2013}.
In this study, a Lucid polarization camera was used to simultaneously acquire the holograms required for parallel phase-shifting holography, and real-time acquisition and reconstruction were implemented in MATLAB. 
The MATLAB code used in this study has also been made openly available as open-source software.
An example of the real-time acquisition is provided as a video in the Supplementary Material.
It should be noted that, because PPLN is a birefringent crystal, the ordinary and extraordinary waves experience different refractive indices. 
Accordingly, the optical path length should be compensated by adjusting the position of the mirror 2 in the first signal path so as to eliminate distinguishability arising from the difference in the arrival times of the two signal fields at the sensor. 
The detailed explanation of adjusting the mirror 2 can be seen in Supplemental section 1.

\subsection{Results}
First, we experimentally verified whether the proposed method could simultaneously acquire holograms with different geometric phase shifts.
Off-axis interference fringes were generated by tilting mirror 2 in the signal path corresponding to the first SPDC process. 
The experimental results are shown in Fig. \ref{fig:QHUP_Fig.3}.
In this evaluation, 1224 $\times$ 1024 pixels were used.
The results confirmed that holograms with different geometric phase shifts were simultaneously acquired as theoretically predicted.
The relative phase differences between the interference fringes were evaluated from the phases of the first-order components in their respective Fourier spectra.
This analysis is based on the Fourier shift theorem (see Supplementary Section 3 for details).
The experiment was repeated 10 times, and the geometric phase shifts obtained in each measurement, together with their mean values, were plotted as the constellation map shown on the right-hand side of Fig. 3.
The constellation map confirms that the relative phase shifts between the holograms recorded at the different pixels of the polarization camera are approximately separated by \(\pi/2\), as theoretically expected.
The measured relative phase differences are summarized in Table \ref{tab:GP} and confirm that phase shifts with approximately $\pi/2$ intervals were successfully achieved. 
The deviations from the ideal phase shifts are attributed to a slight misalignment of the fast axis of the QWP1 experienced by the signal field generated in the first SPDC process.
The experimental results show that the holograms recorded at pixels corresponding to different polarizer orientations exhibit the expected phase shifts, in good agreement with the theoretical prediction.

Then, the aberrations present in the optical system were characterized.
Through post-processing, the phase distribution and the corresponding Zernike coefficients were measured.
The obtained hologram with 1296 $\times$ 1296 pixels was used for the evaluation, and a circular region with a diameter of 1024 pixels was extracted and used for Zernike fitting.
The measured phase distribution was fitted using Zernike polynomials, and the fitting results are shown in Fig. \ref{fig:QHUP_Fig.4}. 
The measured phase distribution was fitted using Zernike polynomials up to the 21st term.
Because the experimentally retrieved phase distribution was affected by noise, a \(16\times16\) spatial smoothing filter was applied. 
In addition, since the phase measured by the interferometer is wrapped within a \(2\pi\) range, phase unwrapping was performed before fitting the phase distribution with the Zernike polynomials.
In this research, the transport-of-intensity equation-based phase-unwrapping technique \cite{Martinez-Carranza:17,Zhao_2019} was used.
The results shown in Fig. \ref{fig:QHUP_Fig.4}(d) indicate that defocus \(Z_4\) and astigmatism \(Z_5\) and \(Z_6\) are the dominant aberrations in the optical system, excluding piston \(Z_1\) and tip \(Z_2\)/tilt \(Z_3\).
By integrating the proposed method with adaptive wavefront-correction techniques \cite{Saita:15,Pozzi:20}, real-time aberration compensation using devices such as deformable mirrors or deformable lenses could be implemented.

\begin{figure*}[t]
\centering
\includegraphics[width=\linewidth]{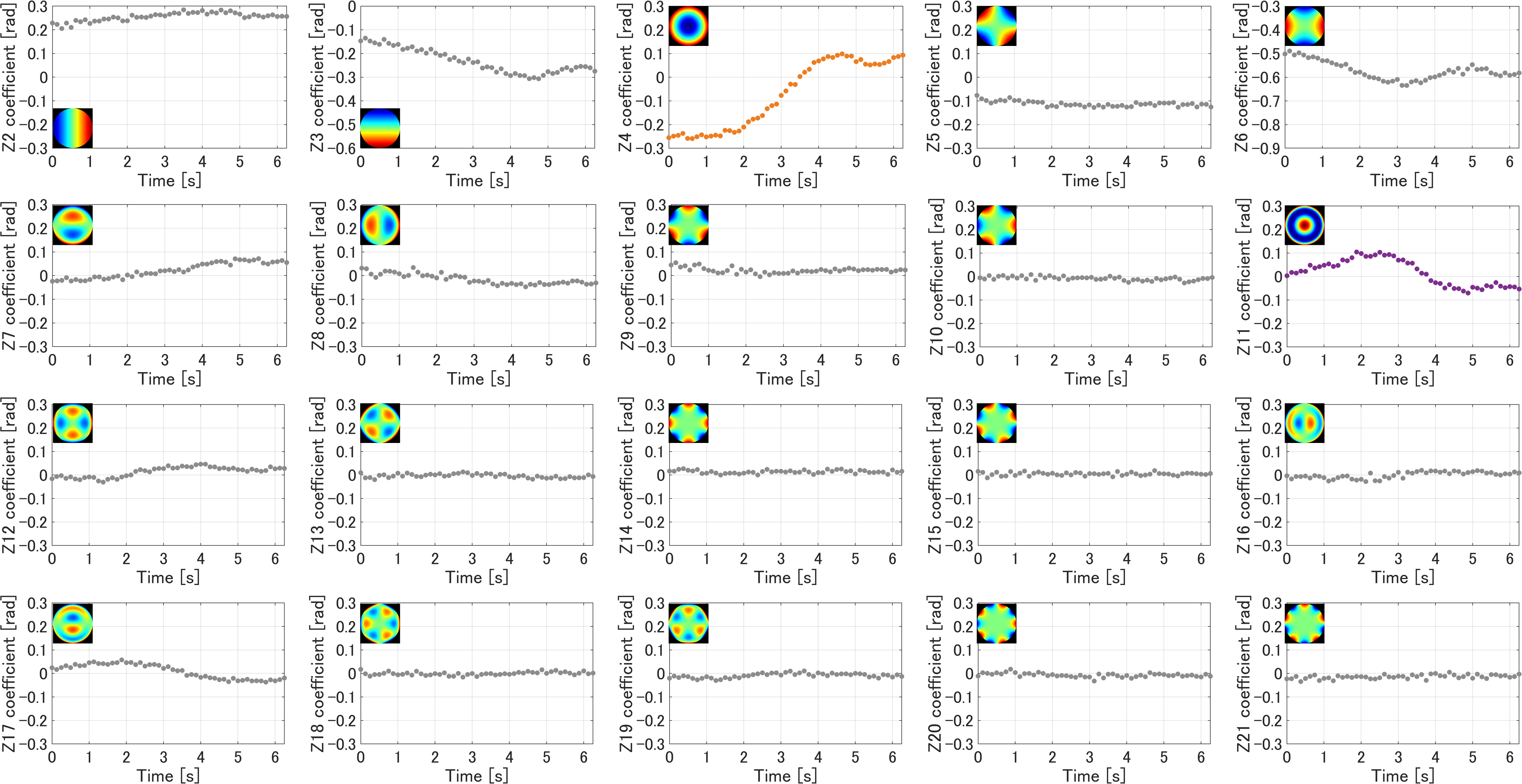}
\caption{Changes in the Zernike coefficients upon applying a voltage to the LC lens.}
\label{fig:QHUP_Fig.6}
\end{figure*}
To demonstrate the real-time capability of the proposed method, the phase modulation produced by a liquid-crystal (LC) lens \cite{Sato_1979,10.1117/12.808123} was observed in real time.
An LC lens is an electrically tunable optical element whose effective curvature can be varied by applying a voltage, which is used in various applications such as classical self-interference holography \cite{Brooker:13,Nobukawa:26,10.1117/1.APN.5.5.056008}. 
A photograph of the LC lens used in the experiment is shown in the upper-right corner of Fig. \ref{fig:QHUP_Fig.2}.
The slow axis of the LC lens was aligned with the polarization direction of the idler photons generated under type-0 phase-matching conditions.
The experimental results are shown in Fig. \ref{fig:QHUP_Fig.5}.
The frame rate and the exposure time were set to 8 fps and 1/8 s, respectively.
The applied voltage to the LC lens was set to 2 V.
The results of Fig. \ref{fig:QHUP_Fig.5}(a) indicate that the proposed method successfully measured the change in lens wavefront as a function of the applied voltage in real time. 
As in the aberration measurement, variations in the Zernike coefficients were quantitatively evaluated by fitting the measured phase distributions with Zernike polynomials in post-processing. 
Figure \ref{fig:QHUP_Fig.5}(b) shows the temporal variation of the phase distribution after Zernike fitting. 
The phase distributions after Zernike fitting without an initial phase distribution shown in Fig. \ref{fig:QHUP_Fig.4}(c) are shown in Fig. \ref{fig:QHUP_Fig.5}(c).
Figure \ref{fig:QHUP_Fig.5}(d) shows the variation of the defocus term alone, demonstrating that the phase modulation introduced by the LC lens was successfully observed.
Figure \ref{fig:QHUP_Fig.5}(e) shows the corresponding changes in the Zernike coefficients.
A video showing the real-time phase measurement, the Zernike-fitted distribution after subtraction of the initial phase, the defocus distribution, and the temporal variation of the Zernike coefficients is provided in the Supplementary Material.
In addition to the defocus coefficient, the spherical-aberration coefficient \(Z_{11}\) also varies with the applied voltage. 
This behavior indicates that, as the voltage applied to the LC lens changes, the curvature of the LC lens gradually increases and higher-order wavefront components beyond the paraxial approximation become observable.
The detailed changes in the Zernike coefficients are shown in Fig. \ref{fig:QHUP_Fig.6}. 
Here, we show the temporal variations of all Zernike terms except piston, which was used as the bias phase for the Zernike polynomial fitting.
The defocus term \(Z_4\) and spherical aberration term \(Z_{11}\), which are associated with changes in the phase profile of the lens, exhibit rapid variations with increasing applied voltage. 
In contrast, the astigmatism term \(Z_5\) shows little change. 
Although the other astigmatism term, \(Z_6\), varies more significantly than \(Z_5\), this behavior is attributed to a slight horizontal tilt of the LC lens in the optical setup.
The coefficients of \(Z_3\), \(Z_7\), and \(Z_{17}\) also change with the applied voltage.
As is evident from Fig. \ref{fig:QHUP_Fig.5}(c), this behavior is attributed to the center of the lens being decentered from the beam center in the vertical direction.
These experimental results demonstrate that voltage-induced changes in the LC lens can be monitored using undetected photons at 1550 nm.
In the field of remote voltage sensing, eye-safe wavelengths are advantageous, and measurements under low-light and low-noise conditions are desirable. 
Therefore, the present results also suggest the potential applicability of the proposed method to remote voltage sensing.

\section{Discussion}
The present study demonstrated real-time QHUP.
However, the measurement signal-to-noise ratio achieved in the present experiment remains lower than that of conventional holography using classical light, primarily because of the low quantum efficiency of the polarization camera at the signal wavelength, around 800 nm.
As demonstrated in the original QHUP study \cite{doi:10.1126/sciadv.abl4301}, the achievable phase accuracy depends on the number of detected photons. 
In the present experimental system, the relatively low pump laser power of 65 mW compared with those used in other QHUP studies, the limited extinction ratio of the wire-grid polarizers integrated into the polarization camera, and the relatively low interference visibility further degrade the measurement accuracy.
These limitations could be mitigated by increasing the available photon flux, for example through the use of induced parametric down-conversion, and by employing a sensor with higher quantum efficiency at the signal wavelength together with a wire-grid polarizer array having a higher extinction ratio.

\section{Conclusion}
In this study, we proposed a single-shot method for acquiring the phase-shifted holograms required for QHUP by using the polarization-filter array of a commercially available polarization camera as a quantum eraser. 
The proposed method exploits geometric phase shifts arising from the different polarization evolutions experienced by the signal fields generated in the first and second SPDC processes, thereby eliminating the need for mechanical mirror scanning for phase shifting.
In addition, the proposed approach offers a larger space-bandwidth product than off-axis holography.
As a proof-of-principle demonstration, we experimentally confirmed that the theoretically predicted geometric phase shifts were correctly encoded in the acquired holograms.
We further demonstrated real-time measurement of optical-system aberrations and dynamic phase changes introduced by a LC lens. 
The experimental results indicate that changes in the Zernike coefficients can be quantitatively monitored.
These results demonstrate the feasibility of real-time acquisition of information carried by undetected photons and suggest potential applications in real-time wavefront sensing and correction, as well as real-time infrared imaging for applications such as greenhouse-gas monitoring.
The optical components and experimental setup used in the proposed method are publicly available as open-source resources, together with the software developed for real-time reconstruction \cite{yoneda_2026_22659491}.
We expect that making these resources openly available will accelerate the practical implementation of QHUP based on the quantum eraser.

\begin{backmatter}
\bmsection{Funding} 
This research is supported by JSPS KAKENHI (25H00854, 26K01415), JST PRESTO (JPMJPR25FA), JST BOOST, (JPMJBS2410), JST SPRING, (JPMJSP2148), and JST CREST (JPMJCR25R1).

\bmsection{Acknowledgment} 
The authors thank Dr. Ayano Tanabe of Citizen Watch Co., Ltd., for supplying the LC lenses. 
The authors gratefully acknowledge Covesion, Hananuma, and Kyocera for providing the 3D CAD models of their components and for their technical support during the development and documentation of the optical setup.

\bmsection{Disclosures} 
The authors declare no conflicts of interest.

\bmsection{Data Availability}
Data underlying the results presented in this paper are available from the corresponding author upon reasonable request.
The three-dimensional design files of the experimental optical system and the complete list of optical components are publicly available in \cite{yoneda_2026_22659491}. 

\bmsection{Code Availability}
The MATLAB code used for real-time quantum holography with undetected photons is available in the GitHub repository https://github.com/NaruYone/QE-QHUP. 
The repository is currently private for limited sharing during manuscript preparation and peer review, and will be made publicly available upon publication. The experimental optical design, including the 3D CAD model, technical drawings, and bill of materials, is openly available through the Zenodo repository \cite{yoneda_2026_22659491} [https://doi.org/10.5281/zenodo.22659491].

\end{backmatter}

\bibliography{sample}

\end{document}


\maketitle

\section{Experiment for quantum eraser in QHUP}
The experimental results shown in Fig. 1(b) were obtained using an image sensor (ORCA-Flash 4.0 C13440-20CU, Hamamatsu Photonics). The result on the right-hand side of Fig. 1(b) was acquired with a polarizer (WP25M-VIS, Thorlabs) placed in front of the sensor.
The other optical components were the same as those described in the main text.
First, QWP1 was placed in the signal path corresponding to the first SPDC process, with its fast axis set at 45 degrees, so that the polarization of the first signal field became orthogonal to that of the second signal field.
Because the PPLN crystal is birefringent, the ordinary and extraordinary waves experience different refractive indices, resulting in a temporal delay between the corresponding polarization components.
The group refractive indices were calculated from the Sellmeier equations, yielding a group-delay difference of approximately 3.1 ps. 
This delay is substantially longer than the coherence time of 0.32 ps, which is determined by the bandwidth, 3 nm, of the band-pass filter placed in front of the image sensor.
Therefore, to erase the temporal distinguishability between the two signal fields, an optical path-length difference of 0.93 mm, corresponding to the product of the group-delay difference and the speed of light, was introduced into the optical path of the signal field generated in the first SPDC process. Because this path has a Michelson-type round-trip configuration, this optical path-length compensation corresponds to a displacement of Mirror 2 by 0.465 mm.

\section{Principle of QHUP}

\subsection{Derivation of the target complex amplitude without considering polarization}

We first derive the principle of quantum holography with undetected photons
(QHUP) without the polarization degree of freedom.
For clarity, we consider the low-gain regime of spontaneous parametric
down-conversion (SPDC), where the probability of generating more than one
signal--idler photon pair is negligible.

When a pump photon is incident on a nonlinear optical crystal, the state after
the first passage through the crystal can be written as
\begin{equation}
|\Psi_{1}\rangle
=
|p\rangle
+
g_{1}e^{i\phi_{p1}}
|s_{1}\rangle
|i_{1}\rangle .
\label{eq:state_first_spdc}
\end{equation}
Here, $|p\rangle$ denotes the component in which the pump photon is not
down-converted, while $|s_{1}\rangle$ and $|i_{1}\rangle$ denote the signal
and idler photons generated in the first SPDC process, respectively.
The coefficient $g_{1}$ denotes the pair-generation amplitude, and
$\phi_{p1}$ represents the phase of the pump field associated with the first
SPDC process.

The propagation of the signal and idler photons through their respective
optical paths is described using complex amplitude transmission coefficients,
\begin{align}
t_{sj}
&=
\sqrt{\eta_{sj}}\,
e^{i\varphi_{sj}},
\label{eq:tsj}
\\
t_{ij}
&=
\sqrt{\eta_{ij}}\,
e^{i\varphi_{ij}},
\label{eq:tij}
\end{align}
where $\eta_{sj}$ and $\eta_{ij}$ denote the intensity transmittances of the
signal and idler paths associated with the $j$th SPDC process, respectively,
and $\varphi_{sj}$ and $\varphi_{ij}$ denote the corresponding propagation
phases.

The complex amplitude transmission coefficient of the object placed in the
idler path is similarly defined as
\begin{equation}
t_o
=
\sqrt{\eta_o}\,
e^{i\phi_o},
\label{eq:object_complex_transmission}
\end{equation}
where $\eta_o$ and $\phi_o$ represent the intensity transmittance and phase of the object, respectively.

The idler photon generated in the first SPDC process interacts with the object.
The transformation of the first idler photon can be expressed as
\begin{equation}
|i_1\rangle
\longrightarrow
t_{i1}t_o|i_1\rangle
+
r_1|l_1\rangle ,
\label{eq:idler1_transformation}
\end{equation}
where $|l_1\rangle$ represents an effective loss mode that includes
absorption, scattering, and coupling to modes that are distinguishable from
the relevant idler mode.
The coefficients satisfy
\begin{align}
|t_{i1}t_o|^2+|r_1|^2
&=
1.
\label{eq:loss_normalization_1}
\end{align}
Here, $r_1$ is an effective loss amplitude including both the propagation loss
in the first idler path and the loss introduced by the object.
Note that loss of an idler photon does not imply loss of its
correlated signal photon.
For example, when the first idler photon is absorbed, scattered, or coupled
into $|l_1\rangle$, its correlated signal photon $|s_1\rangle$ remains in the
system and can still be detected.
The same argument applies to the second SPDC process.
Therefore, the loss-mode components must be retained when calculating the
signal detection probability.
After reflection of the pump and the relevant optical fields, the pump
re-enters the same nonlinear crystal and the second SPDC process becomes
possible.
Taking the propagation and loss of both idler alternatives into account, the
state after the second SPDC process can be written as
\begin{equation}
\begin{aligned}
|\Psi_{2}\rangle
={}&
|p\rangle
+
g_{1}e^{i\phi_{p1}}
t_{s1}t_{i1}t_o
|s_{1}\rangle|i_{1}\rangle
+
g_{1}e^{i\phi_{p1}}
t_{s1}r_{1}
|s_{1}\rangle|l_{1}\rangle
+
g_{2}e^{i\phi_{p2}}
t_{s2}
|s_{2}\rangle|i_{2}\rangle
\\
={}&
|p\rangle
+
g_{1}e^{i\phi_{p1}}
t_{s1}t_{i1}t_o
|s_{1}\rangle|i_{1}\rangle
+
g_{1}e^{i\phi_{p1}}
t_{s1}r_{1}
|s_{1}\rangle|l_{1}\rangle
+
g_{2}e^{i\phi_{p2}}
t_{s2}t_{i2}
|s_{2}\rangle|i_{2}\rangle
+
g_{2}e^{i\phi_{p2}}
t_{s2}r_{2}
|s_{2}\rangle|l_{2}\rangle ,
\label{eq:state_second_spdc_general}
\end{aligned}
\end{equation}
where the idler photon associated with the second SPDC process can be
written as
\begin{equation}
|i_2\rangle
\longrightarrow
t_{i2}|i_2\rangle
+
r_2|l_2\rangle ,
\label{eq:idler2_transformation}
\end{equation}
where $|l_2\rangle$ denotes the corresponding orthogonal loss or unmatched
mode.
The coefficients satisfy
\begin{align}
|t_{i2}|^2+|r_2|^2
&=
1.
\label{eq:loss_normalization_2}
\end{align}
The interference between the two SPDC processes depends on whether the final
states associated with the two possible pair-generation processes are
distinguishable.
We first consider the ideal case in which both the signal and idler photons
generated in the first and second SPDC processes are indistinguishable, $|s_1\rangle = |s_2\rangle = |s\rangle$ and $|i_1\rangle = |i_2\rangle = |i\rangle$.
The common signal and idler modes are assumed to be orthogonal to the loss
modes, and the loss modes associated with the two alternatives are assumed to
be mutually orthogonal, $\langle i|l_1\rangle = \langle i|l_2\rangle = \langle l_1|l_2\rangle = 0$.
Under these conditions, Eq.~(\ref{eq:state_second_spdc_general}) becomes
\begin{equation}
\begin{aligned}
|\Psi_{2}\rangle
={}&
|p\rangle
+
\left(
g_{1}e^{i\phi_{p1}}
t_{s1}t_{i1}t_o
+
g_{2}e^{i\phi_{p2}}
t_{s2}t_{i2}
\right)
|s\rangle|i\rangle
+
g_{1}e^{i\phi_{p1}}
t_{s1}r_1
|s\rangle|l_1\rangle
+
g_{2}e^{i\phi_{p2}}
t_{s2}r_2
|s\rangle|l_2\rangle .
\label{eq:state_second_spdc_indistinguishable}
\end{aligned}
\end{equation}
Equation~(\ref{eq:state_second_spdc_indistinguishable}) explicitly shows
which probability amplitudes can interfere.
The first and second SPDC processes both contain components that terminate in
the same signal--idler state $|s\rangle|i\rangle$.
Therefore, these two probability amplitudes are indistinguishable and can
interfere.
In contrast, the components containing $|l_1\rangle$ and $|l_2\rangle$ are
distinguishable from the common idler mode and from each other, and therefore
do not contribute to the interference term.
Projecting Eq.~(\ref{eq:state_second_spdc_indistinguishable}) onto the detected
signal state gives
\begin{equation}
\begin{aligned}
\langle s|\Psi_{2}\rangle
={}&
\left(
g_{1}e^{i\phi_{p1}}
t_{s1}t_{i1}t_o
+
g_{2}e^{i\phi_{p2}}
t_{s2}t_{i2}
\right)
|i\rangle
+
g_{1}e^{i\phi_{p1}}
t_{s1}r_1
|l_1\rangle
+
g_{2}e^{i\phi_{p2}}
t_{s2}r_2
|l_2\rangle .
\label{eq:conditional_idler_state}
\end{aligned}
\end{equation}
Although the idler photon is not detected, the idler degree of freedom cannot
simply be removed from the state.
Instead, the signal detection probability is obtained by taking the squared
norm of the remaining idler state $P_s
=
\left\|
\langle s|\Psi_{2}\rangle
\right\|^2$.
The probability of the signal photon can be described as
\begin{equation}
\begin{aligned}
P_s
={}&
\left|
g_{1}e^{i\phi_{p1}}
t_{s1}t_{i1}t_o
+
g_{2}e^{i\phi_{p2}}
t_{s2}t_{i2}
\right|^2
+
|g_{1}|^2
|t_{s1}|^2
|r_1|^2
+
|g_{2}|^2
|t_{s2}|^2
|r_2|^2 
\\
={}&
|g_{1}|^2
|t_{s1}|^2
|t_{i1}t_o|^2
+
|g_{1}|^2
|t_{s1}|^2
|r_1|^2
+
|g_{2}|^2
|t_{s2}|^2
|t_{i2}|^2
+
|g_{2}|^2
|t_{s2}|^2
|r_2|^2
\\&+
2\operatorname{Re}
\left[
g_{1}g_{2}^{*}
e^{i(\phi_{p1}-\phi_{p2})}
t_{s1}t_{s2}^{*}
t_{i1}t_{i2}^{*}
t_o
\right].
\label{eq:Ps_before_expansion}
\end{aligned}
\end{equation}
Using Eqs.~(\ref{eq:loss_normalization_1}) and
(\ref{eq:loss_normalization_2}), the first four terms can be simplified as
\begin{equation}
\begin{aligned}
P_s
={}&
|g_{1}|^2|t_{s1}|^2
+
|g_{2}|^2|t_{s2}|^2
+
2\operatorname{Re}
\left[
g_{1}g_{2}^{*}
e^{i(\phi_{p1}-\phi_{p2})}
t_{s1}t_{s2}^{*}
t_{i1}t_{i2}^{*}
t_o
\right]
\\={}&
|g_{1}|^2\eta_{s1}
+
|g_{2}|^2\eta_{s2}
+
2|g_{1}g_{2}|
\sqrt{
\eta_{s1}
\eta_{s2}
\eta_{i1}
\eta_{i2}
\eta_o
}
\cos
\left(
\Delta\Phi_0+\phi_o
\right),
\label{eq:Ps_general_no_pol}
\end{aligned}
\end{equation}
where, taking $g_1$ and $g_2$ as real positive amplitudes, the initial phase
is defined as
\begin{equation}
\Delta\Phi_0
=
(\phi_{p1}-\phi_{p2})
+
(\varphi_{s1}-\varphi_{s2})
+
(\varphi_{i1}-\varphi_{i2}).
\label{eq:initial_phase}
\end{equation}
Equation~(\ref{eq:Ps_general_no_pol}) illustrates an important distinction
between losses in the detected signal paths and those in the undetected idler
paths.
Loss in a signal path directly decreases the corresponding detected signal
intensity through $|t_{s1}|^2$ or $|t_{s2}|^2$.
In contrast, loss in an idler path does not remove the already generated
signal photon.
Instead, it decreases the amplitudes of the idler components that occupy the
common indistinguishable mode $|i\rangle$ and therefore reduces the
interference term.
In this sense, the idler losses are retained in the interference term through
$t_{i1}t_{i2}^{*}$, while they cancel from the individual signal-background
terms owing to the contributions of the corresponding loss modes
$|l_1\rangle$ and $|l_2\rangle$.
When the propagation losses are negligible and the two SPDC amplitudes are equal,
Eq.~(\ref{eq:Ps_general_no_pol}) reduces to
\begin{equation}
P_s
=
2|g|^2
\left[
1+
\sqrt{\eta_o}
\cos
\left(
\Delta\Phi_0+\phi_o
\right)
\right].
\label{eq:Ps_ideal_QHUP}
\end{equation}
Therefore, under the balanced and lossless condition, the object amplitude
$\sqrt{\eta_o}$ directly determines the fringe visibility, whereas the
object phase $\phi_o$ directly shifts the interference phase.
The initial phase $\Delta\Phi_0$ is independent of the object and is
determined by the relative phases of the pump, signal, and idler optical
paths.
It can therefore be measured in advance, or after the object measurement, by
performing a reference measurement without the object.
In conventional QHUP, controlled phase shifting is typically implemented by
changing the optical path length of one of the interferometer arms.
For example, displacement of a mirror in the idler path changes the relative
idler phase
and thereby introduces a controlled phase shift into
Eq.~(\ref{eq:Ps_ideal_QHUP}).
For four-step phase shifting, we denote the four measured signal intensities
by $P_{s,0}$,
$P_{s,\pi/2}$,
$P_{s,\pi}$,
$P_{s,3\pi/2}$.
After subtraction or calibration of the initial phase $\Delta\Phi_0$, the
object phase is obtained as
\begin{equation}
\phi_o
=
\operatorname{atan2}
\left(
P_{s,3\pi/2}-P_{s,\pi/2},
P_{s,0}-P_{s,\pi}
\right),
\label{eq:object_phase_fourstep}
\end{equation}
and the object amplitude is obtained as
\begin{equation}
\sqrt{\eta_o}
=
2
\frac{
\left\{
\left[
P_{s,3\pi/2}-P_{s,\pi/2}
\right]^2
+
\left[
P_{s,0}-P_{s,\pi}
\right]^2
\right\}^{1/2}
}{
P_{s,0}
+
P_{s,\pi/2}
+
P_{s,\pi}
+
P_{s,3\pi/2}
}.
\label{eq:object_amplitude_fourstep}
\end{equation}

\subsection{When distinguishability exists in the signal paths}
The derivation in the previous subsection assumes that the signal photons
generated in the first and second SPDC processes are indistinguishable.
We next consider the case in which the two signal states are distinguishable.
We assume that the idler modes that contribute to the interference are
indistinguishable, $|i_1\rangle=|i_2\rangle=|i\rangle$, whereas the signal states satisfy $\langle s_2|s_1\rangle=0$.
The detector is assumed to detect both signal modes without resolving which
SPDC process generated the detected photon.
The corresponding projection operator can be written as
\begin{equation}
\Pi_s
=
\left(
|s_1\rangle\langle s_1|
+
|s_2\rangle\langle s_2|
\right)
\otimes I_i,
\label{eq:signal_detection_projector}
\end{equation}
where $I_i=
|i\rangle\langle i|
+
|l_1\rangle\langle l_1|
+
|l_2\rangle\langle l_2|$ denotes the identity operator for the undetected idler degrees of
freedom, including the common idler mode and the loss modes.
Using Eq.~(\ref{eq:state_second_spdc_general}), the signal detection
probability is
\begin{equation}
P_s
=
\langle\Psi_2|\Pi_s|\Psi_2\rangle.
\end{equation}
Because $\langle s_2|s_1\rangle=0$, the cross term between the first and
second SPDC processes vanishes.
Using Eqs.~(\ref{eq:loss_normalization_1}) and
(\ref{eq:loss_normalization_2}), the signal detection probability therefore
becomes
\begin{equation}
P_s
=
|g_1|^2\eta_{s1}
+
|g_2|^2\eta_{s2}.
\label{eq:Ps_distinguishable}
\end{equation}
Thus, neither the amplitude $\sqrt{\eta_o}$ nor the phase $\phi_o$ of the
object appears in the detected signal intensity when the two signal states
are completely distinguishable.
More generally, the interference term contains the inner products of the
states associated with the two possible pair-generation processes.
Therefore, interference requires indistinguishability in both the detected
signal degree of freedom and the undetected idler degree of freedom.
In the present experiment, distinguishability is intentionally introduced in
the polarization degree of freedom of the signal photon.

\subsection{Polarization-induced distinguishability and quantum erasure}

We now explicitly include the polarization degree of freedom of the signal
photon.
The spatial, temporal, and spectral modes of the two signal photons are
assumed to be identical, and the signal states are written as
\begin{align}
|s_1\rangle
&\rightarrow
|s\rangle|\epsilon_{s1}\rangle,
\\
|s_2\rangle
&\rightarrow
|s\rangle|\epsilon_{s2}\rangle,
\end{align}
where $|\epsilon_{s1}\rangle$ and $|\epsilon_{s2}\rangle$ denote the
polarization states of the signal photons generated in the first and second
SPDC processes, respectively.

When the two polarization states are orthogonal, $\langle\epsilon_{s2}|\epsilon_{s1}\rangle=0$, the two SPDC processes can be distinguished from the signal polarization.
Consequently, the interference term vanishes even when all other signal and
idler modes are indistinguishable.

To erase the polarization-based distinguishability, a linear polarization
analyzer is placed in front of the signal detector.
The polarization state transmitted by an analyzer oriented at an angle
$\theta$ is
\begin{equation}
|p_\theta\rangle
=
\cos\theta|H\rangle
+
\sin\theta|V\rangle.
\label{eq:analyzer_state}
\end{equation}
The probability amplitudes for the signal photons generated in the two SPDC
processes to pass through the analyzer are defined as
\begin{align}
c_1(\theta)
&=
\langle p_\theta|\epsilon_{s1}\rangle,
\\
c_2(\theta)
&=
\langle p_\theta|\epsilon_{s2}\rangle.
\label{eq:analyzer_coefficients}
\end{align}

After transmission through the analyzer, both signal photons are projected
onto the same polarization state $|p_\theta\rangle$.
Thus, the polarization information that distinguishes the two SPDC processes
is erased.
Using the state introduced in the previous subsection, projection of the
signal polarization onto $|p_\theta\rangle$ gives the remaining idler state
\begin{equation}
\begin{aligned}
|\psi_D(\theta)\rangle_i
={}&
\left[
g_1e^{i\phi_{p1}}
t_{s1}t_{i1}t_o
c_1(\theta)
+
g_2e^{i\phi_{p2}}
t_{s2}t_{i2}
c_2(\theta)
\right]
|i\rangle
\\
&+ 
g_1e^{i\phi_{p1}}
t_{s1}r_1c_1(\theta)
|l_1\rangle +
g_2e^{i\phi_{p2}}
t_{s2}r_2c_2(\theta)
|l_2\rangle.
\label{eq:conditional_idler_polarization}
\end{aligned}
\end{equation}
The signal detection probability is obtained from $P_s(\theta) = \langle\psi_D(\theta)|\psi_D(\theta)\rangle$.
Using the orthogonality of $|i\rangle$, $|l_1\rangle$, and $|l_2\rangle$,
together with Eqs.~(\ref{eq:loss_normalization_1}) and
(\ref{eq:loss_normalization_2}), we obtain
\begin{equation}
\begin{aligned}
P_s(\theta)
={}&
|g_1|^2\eta_{s1}|c_1(\theta)|^2
+
|g_2|^2\eta_{s2}|c_2(\theta)|^2
+
2\operatorname{Re}
\left[
g_1g_2^*
e^{i(\phi_{p1}-\phi_{p2})}
t_{s1}t_{s2}^*
t_{i1}t_{i2}^*
t_o
c_1(\theta)c_2^*(\theta)
\right].
\label{eq:Ps_general_polarization}
\end{aligned}
\end{equation}
The polarization-dependent complex coefficient can be written as
\begin{equation}
c_1(\theta)c_2^*(\theta)
=
|c_1(\theta)c_2^*(\theta)|
e^{i\phi_{\mathrm{pol}}(\theta)},
\label{eq:polarization_cross_term}
\end{equation}
where $\phi_{\mathrm{pol}}(\theta)$ represents the relative phase introduced
by the polarization transformation and projection.
Using the definition of $\Delta\Phi_0$ in
Eq.~(\ref{eq:initial_phase}), Eq.~(\ref{eq:Ps_general_polarization}) becomes
\begin{equation}
\begin{aligned}
P_s(\theta)
={}&
|g_1|^2\eta_{s1}|c_1(\theta)|^2
+
|g_2|^2\eta_{s2}|c_2(\theta)|^2
\\
&+
2|g_1g_2|
\sqrt{
\eta_{s1}\eta_{s2}
\eta_{i1}\eta_{i2}
\eta_o
}
|c_1(\theta)c_2^*(\theta)|
\cos
\left[
\Delta\Phi_0
+
\phi_o
+
\phi_{\mathrm{pol}}(\theta)
\right].
\label{eq:Ps_polarization_compact}
\end{aligned}
\end{equation}
Thus, the polarization projection modifies both the amplitude and phase of the
interference term.
The magnitude $|c_1c_2^*|$ determines the polarization-dependent modulation
of the interference amplitude, whereas $\phi_{\mathrm{pol}}$ introduces an
additional phase shift into the interference fringes.

\subsection{Polarization transformation by a wave plate}

The polarization transformation produced by a linear retarder with fast-axis
orientation $\alpha$ and retardance $\delta$ is described by the Jones
operator
\begin{equation}
Q(\alpha,\delta)
=
R(-\alpha)
\begin{pmatrix}
1&0\\
0&e^{i\delta}
\end{pmatrix}
R(\alpha),
\label{eq:general_retarder}
\end{equation}
where
\begin{equation}
R(\alpha)
=
\begin{pmatrix}
\cos\alpha&-\sin\alpha\\
\sin\alpha&\cos\alpha
\end{pmatrix}.
\end{equation}
For an ideal quarter-wave plate (QWP), $\delta=\frac{\pi}{2}$.
If the initial signal polarization is
$|\epsilon_s^{(0)}\rangle$, the polarization states immediately before the
analyzer can generally be written as
\begin{align}
|\epsilon_{s1}\rangle
&=
U_{s1}|\epsilon_s^{(0)}\rangle,
\\
|\epsilon_{s2}\rangle
&=
U_{s2}|\epsilon_s^{(0)}\rangle,
\end{align}
where $U_{s1}$ and $U_{s2}$ represent the complete polarization
transformations experienced by the signal photons generated in the first and
second SPDC processes, respectively.
The analyzer amplitudes are therefore
\begin{align}
c_1(\theta)
&=
\langle p_\theta|
U_{s1}
|\epsilon_s^{(0)}\rangle,
\\
c_2(\theta)
&=
\langle p_\theta|
U_{s2}
|\epsilon_s^{(0)}\rangle.
\label{eq:general_analyzer_coefficients}
\end{align}
\subsection{Geometric phase shift in the present optical configuration}
We now apply the above formulation to the polarization configuration used in
the present experiment.
The signal photon generated by SPDC is initially vertically polarized,
\begin{equation}
|\epsilon_s^{(0)}\rangle
=
|V\rangle
=
\begin{pmatrix}
0\\
1
\end{pmatrix}.
\end{equation}
The signal photon generated in the first SPDC process experiences the
polarization sequence
\begin{equation}
|V\rangle
\rightarrow
Q_a
\rightarrow
M
\rightarrow
Q_a
\rightarrow
Q_b
\rightarrow
P(\theta),
\label{eq:first_signal_sequence}
\end{equation}
whereas the signal photon generated in the second SPDC process experiences
\begin{equation}
|V\rangle
\rightarrow
Q_b
\rightarrow
P(\theta).
\label{eq:second_signal_sequence}
\end{equation}
Here, $Q_a$ and $M$ denote the QWP traversed twice only by the signal photon generated
in the first SPDC process and mirror 2, whereas $Q_b$ denotes the QWP experienced by the signal photons generated in both SPDC processes.
Assuming that the mirror does not introduce a polarization-dependent relative
phase, its scalar phase can be included in $\Delta\Phi_0$.
The polarization transformations can then be written as
\begin{align}
U_{s1}
&=
Q_bQ_a^2,
\label{eq:specific_polarization_transformations1}
\\
U_{s2}
&=
Q_b.
\label{eq:specific_polarization_transformations2}
\end{align}
For ideal QWPs,
\begin{equation}
\alpha_a
=
\alpha_b
=
\frac{\pi}{4},
\qquad
\delta_a
=
\delta_b
=
\frac{\pi}{2},
\end{equation}
so Eq. (\ref{eq:general_retarder}) can be described as
\begin{equation}
Q_a
=
Q_b
=
Q\left(\frac{\pi}{4},\frac{\pi}{2}\right)
=
\frac{1}{2}
\begin{pmatrix}
1+i & -1+i \\
-1+i & 1+i
\end{pmatrix}
.
\label{eq:general_retardercor}
\end{equation}
Using Eq. (\ref{eq:general_retardercor}), Eq. (\ref{eq:specific_polarization_transformations1}) and Eq. (\ref{eq:specific_polarization_transformations2}) are described as
\begin{align}
U_{s1}
&=
Q_bQ_a^2
=
\frac{1}{2}
\begin{pmatrix}
1-i & -1-i\\
-1-i & 1-i
\end{pmatrix},
\label{eq:specific_polarization_transformations11}
\\
U_{s2}
&=
Q_b
=
\frac{1}{2}
\begin{pmatrix}
1+i & -1+i\\
-1+i & 1+i
\end{pmatrix}.
\label{eq:specific_polarization_transformations22}
\end{align}
\begin{figure}[b]
\centering
\includegraphics[width=100mm]{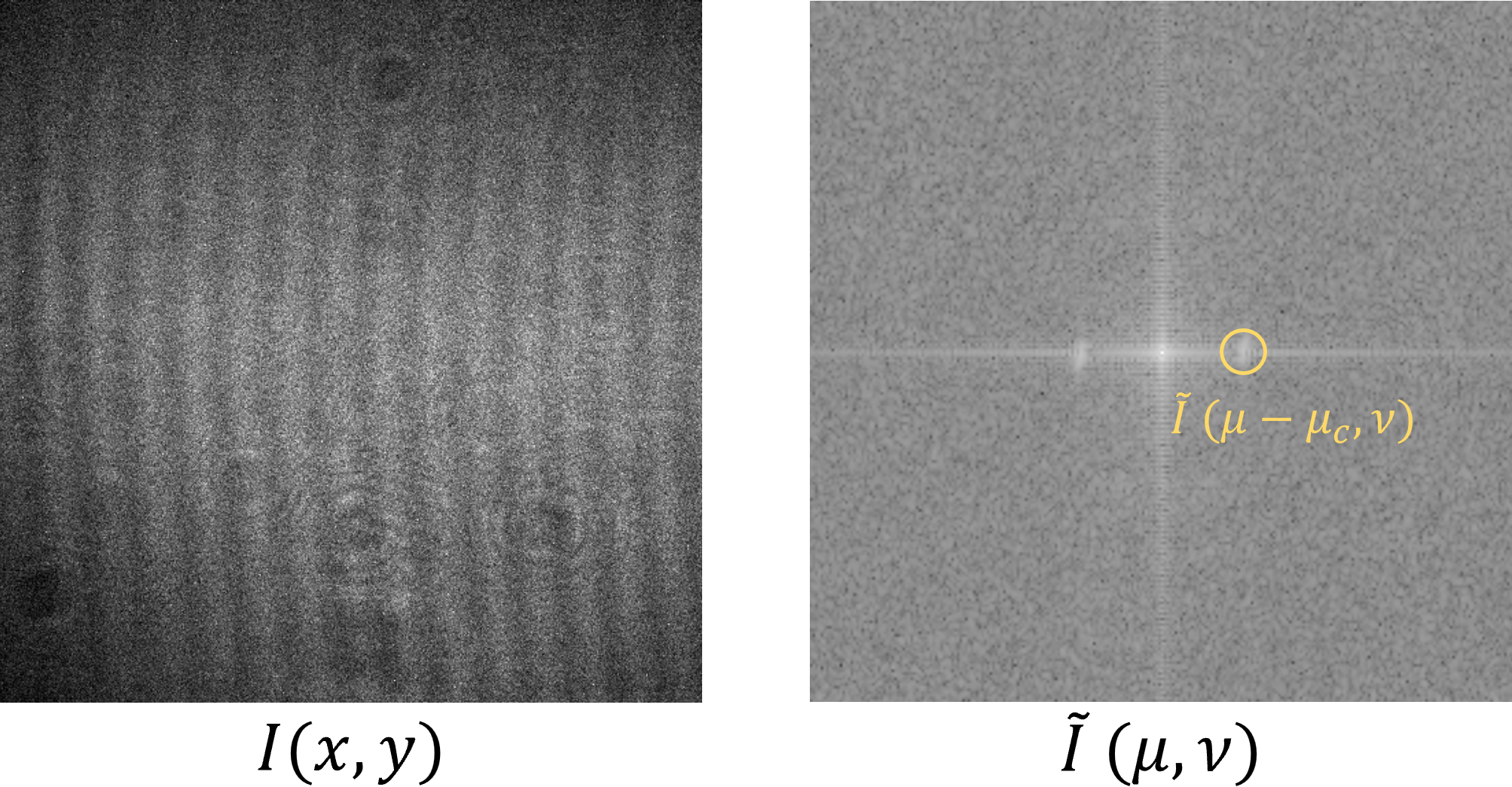}
\caption{Schematic illustration of the geometric phase-shift measurement. The geometric phase shift is obtained from the phase of the Fourier spectrum of the interference fringes.
}
\label{fig:QHUP_Fig.S1}
\end{figure}
The polarization states immediately before the analyzer are then
\begin{align}
|\epsilon_{s1}\rangle
&=
\frac{1}{2}
\begin{pmatrix}
-1-i\\
1-i
\end{pmatrix},
\\
|\epsilon_{s2}\rangle
&=
\frac{1}{2}
\begin{pmatrix}
-1+i\\
1+i
\end{pmatrix}.
\label{eq:ideal_signal_polarizations}
\end{align}
These states satisfy $\langle\epsilon_{s2}|\epsilon_{s1}\rangle=0$ 
and correspond to orthogonal circular polarizations, apart from global phase
factors.
Thus, without the polarization analyzer, the first and second SPDC processes
are distinguishable through the signal polarization and no interference is
observed.
Projection onto the analyzer state in Eq.~(\ref{eq:analyzer_state}) gives
\begin{align}
c_1(\theta)
&=
\frac{1-i}{2}
\left(
\sin\theta-i\cos\theta
\right),
\\
c_2(\theta)
&=
\frac{1-i}{2}
\left(
-\cos\theta+i\sin\theta
\right).
\label{eq:c1c2_ideal}
\end{align}
Their magnitudes are
\begin{equation}
|c_1(\theta)|^2
=
|c_2(\theta)|^2
=
\frac{1}{2},
\end{equation}
and their complex cross term is
\begin{equation}
c_1(\theta)c_2^*(\theta)
=
\frac{1}{2}
e^{i(2\theta+\pi/2)}.
\label{eq:ideal_polarization_cross_term}
\end{equation}
Therefore,
\begin{equation}
\phi_{\mathrm{pol}}(\theta)
=
2\theta+\frac{\pi}{2}.
\label{eq:ideal_polarization_phase}
\end{equation}
The constant phase $\pi/2$ depends on the Jones-matrix convention and can be
included in the initial phase calibration.
Substituting Eq.~(\ref{eq:ideal_polarization_cross_term}) into
Eq.~(\ref{eq:Ps_polarization_compact}) gives
\begin{equation}
\begin{aligned}
P_s(\theta)
={}&
\frac{1}{2}
\left(
|g_1|^2\eta_{s1}
+
|g_2|^2\eta_{s2}
\right)
\\
&+
|g_1g_2|
\sqrt{
\eta_{s1}\eta_{s2}
\eta_{i1}\eta_{i2}
\eta_o
}
\cos
\left(
\Delta\Phi_0
+
\phi_o
+
2\theta
+
\frac{\pi}{2}
\right).
\label{eq:Ps_geometric_phase_general}
\end{aligned}
\end{equation}
Under balanced and lossless conditions, by neglecting bias phases $\Delta\Phi_0$ and $\frac{\pi}{2}$,
Eq.~(\ref{eq:Ps_geometric_phase_general}) reduces to
\begin{equation}
P_s(\theta)
=
|g|^2
\left[
1+
\sqrt{\eta_o}
\cos
\left(
\phi_o
+
2\theta
\right)
\right].
\label{eq:Ps_geometric_phase_ideal}
\end{equation}
For analyzer orientations $\theta$ set to $
0$ $\frac{\pi}{4}$, $
\frac{\pi}{2}$, $\frac{3\pi}{4}$, the corresponding geometric phases are $0$, $\frac{\pi}{2}$, $\pi$, and $\frac{3\pi}{2}$, respectively.
Thus, the four analyzer orientations introduce relative phase shifts separated
by $\pi/2$.
The object amplitude and phase can therefore be reconstructed using the same
four-step phase-shifting procedure.

\section{Geometric phase measurements from Fourier spectrum}
The phase shift of the interference fringes can be quantitatively evaluated from the phase of the first-order diffraction component in the Fourier domain. 
By tilting a mirror in a path of the first signal photon, the observed interference fringe distribution is expressed as

\begin{equation}
I(x,y)\propto 1+\cos\left(2\pi \mu_c x
+
2\theta
\right),
\end{equation}
where $\mu_c$ is the spatial carrier frequency of the interference fringe. 
Using Euler's formula, Eq.~(1) can be rewritten as

\begin{equation}
I(x,y)
\propto
1+
\frac{1}{2}e^{i
2\theta
}e^{i2\pi \mu_c x}
+
\frac{1}{2}e^{-i
2\theta
}e^{-i2\pi \mu_c x}.
\end{equation}
After Fourier transformation, the zeroth- and first-order diffraction components are spatially separated in the Fourier domain. 
The Fourier spectrum can be described as
\begin{equation}
\tilde{I}(\mu,\nu)
\propto
\delta(\mu,\nu)
+
\frac{1}{2}e^{i2\theta
}\delta(\mu - \mu_c,\nu)
+
\frac{1}{2}e^{-i2\theta
}\delta(\mu + \mu_c,\nu).
\end{equation}
The positive first-order component is given by
\begin{equation}
\tilde{I}(\mu - \mu_c,\nu)
\propto
\frac{1}{2}e^{i2\theta
}.
\end{equation}
Therefore, the geometric phase can be directly obtained from the argument of the positive first-order diffraction component.
